\pdfoutput=1
\documentclass[aip,jap,reprint,amsmath,amssymb,floatfix]{revtex4-2}

\usepackage{graphicx}
\usepackage{bm}
\usepackage{amsfonts}
\usepackage{booktabs}
\usepackage{tikz}
\usetikzlibrary{positioning, arrows.meta}
\usepackage[colorlinks=true,linkcolor=blue,citecolor=blue,urlcolor=blue]{hyperref}

\begin{document}

\title{Topological Self-Healing Physical Intelligence in Continuous Ginzburg--Landau Fields}

\author{Takashi Hikihara}
\email[Author to whom correspondence should be addressed: ]{hikihara.takashi.2n@kyoto-u.ac.jp, thikihara@nii.ac.jp}
\affiliation{Professor Emeritus, Kyoto University, Yoshida-honmachi, Sakyo, Kyoto 606-8501, Japan}
\affiliation{Project Professor, National Institute of Informatics, 2-1-2 Hitotsubashi, Chiyoda-ku, Tokyo 101-8430, Japan}

\date{\today}

\begin{abstract}
Ensuring resilience at the physical layer against data anomalies without introducing software-level algorithmic latency remains a critical challenge in non-equilibrium physical computing. Here, we demonstrate that a continuous macroscopic quantum field governed by the time-dependent Ginzburg--Landau (TDGL) equations exhibits autonomous functional self-healing driven by dynamic topological defect nucleation. By breaking spatial inversion symmetry in a 50-fluxoid superconducting ring crystal through non-reciprocal advection flow, the order parameter field self-organizes into a low-dimensional topological manifold characterized by an oscillatory, spatiotemporally circulating memory profile. Under severe chaotic over-injection, the internal viscosity of the medium establishes a passive incubation buffer before triggering a localized topological phase transition. This transition decouples a central vortex pair that acts as a hardware-native physical interlock, sequestering excess kinetic energy and protecting peripheral functional domains. Once the perturbation ends, the system monitors its macroscopic thermodynamic dissipation velocity as an autonomous trigger for online readout recalibration, reducing the inference mean squared error by over two orders of magnitude to match the post-repair geometry. Supported by concrete nanofabrication blueprints in thin-film niobium and theoretical extensions to photonic and cold-atom platforms, this work establishes a universal physical framework for topological material intelligence in open dissipative systems.
\end{abstract}

\maketitle

\section{Introduction}

Physical reservoir computing (PRC) has emerged as an energy-efficient framework for edge computing and temporal data processing by exploiting the nonlinear spatiotemporal dynamics of physical substrates.\cite{jaeger2001, tanaka2019} By mapping high-dimensional computations onto the natural evolution of solid-state devices, optical cavities, or mechanical resonators, PRC bypasses the significant training overhead required by recurrent neural networks. However, deploying these continuous physical systems in non-stationary edge environments reveals a fundamental vulnerability: establishing operational reliability against temporal data anomalies and environmental disruptions remains difficult. When a continuous computational medium experiences severe localized stress or out-of-distribution input streams, the injected kinetic energy distorts the internal phase space, causing cumulative inference degradation and eventual synchronization failure.

In existing implementations, the physical medium operates primarily as a passive dynamical transform. Consequently, fault-mitigation strategies have relied almost exclusively on external digital algorithms or auxiliary diagnostic circuits to detect and compensate for computational errors. This separation of analog processing and digital monitoring introduces notable latency, computational overhead, and circuit complexity, countering the zero-latency objective of physical coprocessors. To date, whether an analog computational substrate can autonomously mitigate disruptions and recover its functional capacity---without digital supervision---has remained largely unexplored in physical computing.

Here, we show that autonomous functional self-healing can be realized natively within a computing substrate through the interplay of intrinsic material dissipation and an engineered gradient flow. We examine the continuous macroscopic quantum field of a superconducting vortex crystal governed by the time-dependent Ginzburg--Landau (TDGL) equations.\cite{ginzburg1950, tinkham1975} Rather than relying on discrete nodal networks, our platform uses the off-diagonal long-range order\cite{yang1962} of a 50-fluxoid ring crystal confined within a circular nanochannel. By breaking the spatial inversion symmetry of the isotropic quantum diffusion field with a directional transport vector,\cite{hikihara2013} we produce an irreversible, advection-dominated gradient flow. This non-reciprocal structure establishes a causal pipeline across the continuous medium, projecting input streams onto a stable topological manifold.

In this geometry, structural self-repair emerges as an intrinsic thermodynamic response of the continuous field dynamics. When an intense, localized chaotic input over-injects energy into the gradient flow, the medium's internal viscosity serves as a passive incubation buffer. When the accumulated phase strain exceeds the local Ginzburg--Landau potential barrier, the lattice undergoes a localized topological phase transition\cite{hikihara1995, machida1995} rather than a global collapse. This transition nucleates and decouples a central vortex pair that functions as a physical interlock,\cite{moon1994} sequestering excess kinetic energy and preserving the integrity of peripheral computing nodes. Once the perturbation ends, the subsequent thermodynamic relaxation provides an autonomous physical trigger to recalibrate the linear readout, reducing the inference mean squared error (MSE) by over two orders of magnitude to match the post-repair geometry. These findings outline a physical approach to resilient material intelligence operating within macroscopic quantum regimes.

\section{Theoretical Model and Numerical Methods}

\subsection{Continuous TDGL Causal Flow and Gauge-Invariant Symmetry Breaking}

To establish a continuous physical medium capable of autonomous memory retention and self-repair, we utilize the spatiotemporal evolution of a macroscopic superconducting vortex field governed by the TDGL equations. We define the complex order parameter field as $\psi(x,y) = |\psi|e^{i\theta}$, where $\theta$ represents the macro-quantum phase. The time evolution of $\psi$ is coupled to the electromagnetic gauge field defined by the vector potential $\bm{A}(x,y)$. The normalized equation of motion for the order parameter is given by\cite{ginzburg1950, kato1991}
\begin{equation}
    \eta\frac{\partial\psi}{\partial t} = -\left(\frac{\nabla}{i} - \bm{A}\right)^2\psi + \psi\left(1 - |\psi|^2\right),
    \label{eq:tdgl}
\end{equation}
where $\eta$ denotes the viscous dissipation coefficient of the superconducting condensate.

In an unconstrained isotropic bulk superconductor, quantized fluxoids organize into a triangular Abrikosov lattice governed by symmetric London repulsion. Such isotropic diffusion lacks the directional causality required for sequential information processing. To construct a directional computing manifold, we confine 50 quantized fluxoids within a narrow circular tracking channel and break spatial inversion symmetry.

This spatial anisotropy is introduced via a directional transport vector field $\bm{v}_{\mathrm{bias}}$ coupled to the gauge field. The modified covariant derivative operator enforces non-reciprocal spatial perturbation dynamics:\cite{hikihara2013}
\begin{equation}
    \nabla_{\bm{A}} \rightarrow \nabla - i\bm{A} - i\bm{v}_{\mathrm{bias}}.
    \label{eq:covariant}
\end{equation}
By coupling directly with the spatial derivative operator, $\bm{v}_{\mathrm{bias}}$ converts the isotropic quantum diffusion field into an irreversible, advection-dominated gradient flow. The forward transport rate along the azimuthal trajectory substantially exceeds the backward relaxation rate. This non-reciprocity establishes a deterministic causal chain across the topological medium: the forward path integrates the statistical variations of complex time-series inputs, while the weaker reverse path provides a restoring force that dampens common-mode fluctuations.

The operational stability and resilience of this continuous reservoir are governed by the local extrema of the Ginzburg--Landau free energy functional. In a localized sector of the ring channel, structural equilibrium is defined by the minimum of the effective potential density $U_{\mathrm{eff}}$:
\begin{equation}
    U_{\mathrm{eff}}(\psi) = \frac{1}{2}\alpha|\psi|^2 + \frac{1}{4}\beta|\psi|^4 + \left|\left(\nabla - i\bm{A} - i\bm{v}_{\mathrm{bias}}\right)\psi\right|^2.
    \label{eq:ueff}
\end{equation}
Under typical computing conditions, the local potential wells remain stable, preserving an equidistant 50-vortex ring configuration. When anomalous kinetic energy stresses this gradient flow, the field exhibits a localized thermodynamic response that initiates structural self-repair.

\subsection{Three-Layer Mathematical Reduction from Discrete Non-Reciprocal Mechanics}

\begin{figure*}[htbp]
\centering
\begin{tikzpicture}[
    layerbox/.style={
        rectangle,
        draw=black,
        thick,
        fill=gray!6,
        rounded corners=2pt,
        text width=0.80\textwidth,
        align=center,
        inner sep=7pt,
        font=\normalsize
    },
    arrow/.style={
        -{Stealth[length=3mm]},
        thick,
        line width=1.1pt
    },
    arrowlabel/.style={
        font=\small\itshape,
        align=center,
        fill=white,
        inner sep=2pt
    }
]

\node[layerbox] (layer1) {\textbf{Layer 1: Discrete Dynamics} (Coupled Real Duffing Lattice)};
\node[layerbox, below=1.0cm of layer1] (layer2) {\textbf{Layer 2: Continuum Field Principle} (Complex TDGL Equation)};
\node[layerbox, below=1.0cm of layer2] (layer3) {\textbf{Layer 3: Physical Implementation} (Engineered Niobium Thin Film)};

\draw[arrow] (layer1) -- (layer2) 
    node[midway, arrowlabel] {Spatial Long-Wavelength Limit ($a \rightarrow 0$) \& Overdamped Limit};
\draw[arrow] (layer2) -- (layer3) 
    node[midway, arrowlabel] {Gauge Invariance \& Microscopic Peierls Substitution ($\phi = 63^\circ$)};

\end{tikzpicture}
\caption{Three-layer hierarchical mapping and physical parameter synchronization pipeline connecting discrete mechanical lattices to continuous macroscopic quantum fields and thin-film realization. Layer 1 defines the baseline discrete network of real-valued, diffusively coupled Duffing oscillators with double-well potentials. Layer 2 represents the emergent macroscale topological continuum governed by the complex TDGL equation, derived via a multiscale perturbation expansion under the spatial long-wavelength limit ($a \rightarrow 0$). Layer 3 details the physical implementation within a high-purity niobium (Nb) thin film ($\xi \approx 40\text{ nm}$). Directional transport is parameterized through a gauge-invariant Peierls substitution with a base coupling of $k_0 = 0.50$ and a curvature-corrected phase twist of $\phi = 63^\circ$ (1.0982 rad), corresponding experimentally to an external perpendicular magnetic field ($39.7\text{ mT}$) or an azimuthal bias current ($150\ \mu\text{A}$).}
\label{fig:layer_mapping}
\end{figure*}
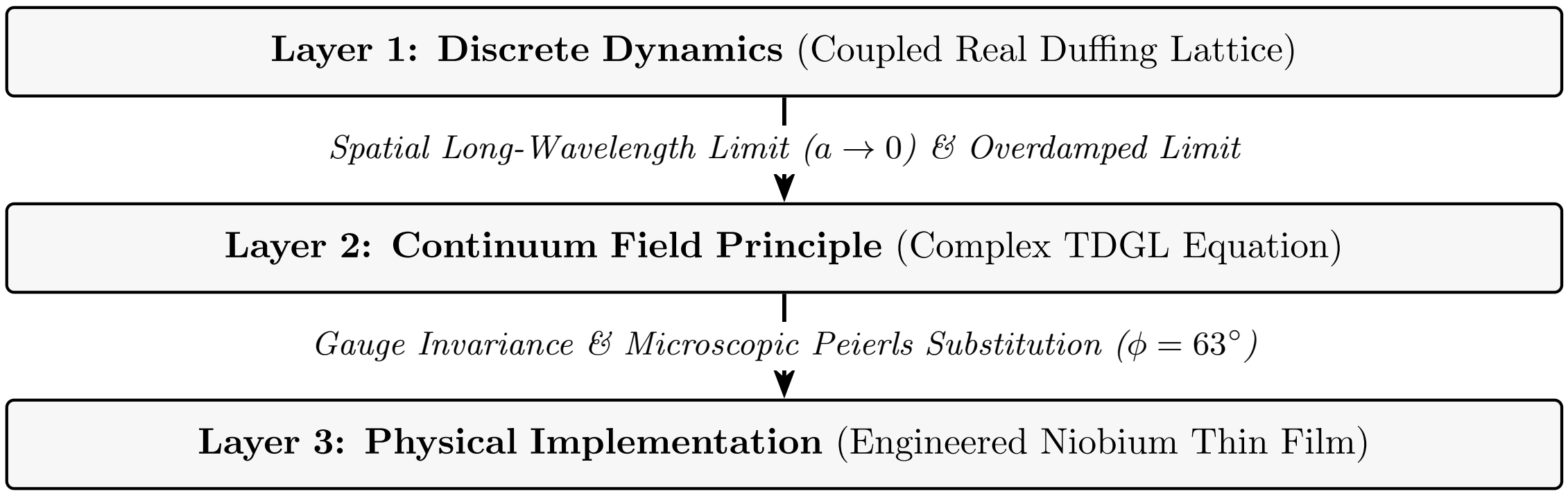

To confirm that the observed self-healing is a universal property of non-reciprocal gradient flows rather than a simulation artifact, we formulate a three-tier hierarchical bridge linking discrete nonlinear mechanics to physical quantum hardware (Fig.~\ref{fig:layer_mapping}).

\textit{Layer 1 to Layer 2 (Mathematical Continuum Limit):} As depicted in Fig.~\ref{fig:layer_mapping}, Layer 1 consists of a discrete dynamical lattice of real-valued, diffusively coupled Duffing oscillators possessing local double-well potentials.\cite{hikihara2026} By defining a slowly varying complex envelope field that accounts for both amplitude variations and inter-node phase differences, we perform a multiscale perturbation expansion (Supplementary Note 1). In the spatial long-wavelength limit ($a \rightarrow 0$) and temporal overdamped limit, the discrete lattice converges rigorously to the continuous complex TDGL equation (Layer 2). Unlike standard empirical phase reductions that neglect amplitude variations, this framework retains the full spatiotemporal envelope degrees of freedom, ensuring that topological defect nucleation---the core of the self-healing response---emerges naturally from the continuum limit.

\textit{Layer 2 to Layer 3 (Physical Nanotechnology Mapping):} The continuum equation is mapped onto a physical thin-film platform (Layer 3) by applying structural and material boundary conditions (Supplementary Note 2). In an engineered $150\text{-nm}$-wide circular track fabricated from high-purity niobium (Nb) ($\xi \approx 40\text{ nm}$), channel curvature induces an azimuthal shear strain of $+3^\circ$ relative to the flat-space triangular lattice symmetry ($60^\circ$). To stabilize the 50-vortex configuration within this track, we introduce a curvature-corrected Peierls phase twist threshold of $\phi = 63^\circ$ ($1.0982\text{ rad}$).\cite{hofstadter1976} Setting the base coupling strength to $k_0 = 0.50$ and solving the transcendental relation linking the discrete diffusion and advection coefficients yields the numerical coupling constants $k_{\mathrm{fwd}} = 0.45$ and $k_{\mathrm{rev}} = 0.005$. These advection rates match standard laboratory operating parameters: an external perpendicular magnetic field of $39.7\text{ mT}$ or a steady azimuthal transport current of $150\ \mu\text{A}$.

\subsection{Numerical Discretization, Manifold Extraction, and Dissipation Trigger}

The spatiotemporal evolution of the order parameter $\psi(x,y)$ and vector potential $\bm{A}(x,y)$ was computed on an $83 \times 83$ spatial grid ($\Delta L \approx 28.77\text{ nm}$, corresponding to an $81 \times 81$ interior computational domain) using high-order finite differences.\cite{kato1991, miyamoto2004} Temporal integration employed a fourth-order Runge--Kutta scheme with a time step bounded by the Courant--Friedrichs--Lewy (CFL) stability criterion. Confinement walls and artificial pinning boundaries were enforced by setting $|\psi|^2 = 0$, while non-reciprocal advection was maintained via the constant azimuthal transport vector $\bm{v}_{\mathrm{bias}}$ in Eq.~(\ref{eq:covariant}).

To evaluate computational dimensionality and avoid overfitting from finite observation windows, the internal reservoir state was decoded from the continuous magnetic flux density field $B(x,y) = (\nabla \times \bm{A})_z$\cite{miyamoto2004} across all $M = 6,561$ virtual grid nodes, forming a spatiotemporal state matrix $\bm{X} \in \mathbb{R}^{T_{\mathrm{all}} \times M}$ ($T_{\mathrm{all}} = 41$). Principal component analysis (PCA) was conducted via singular value decomposition of the covariance matrix of $\bm{X}$.\cite{fujii2017, caremel2024} Readout weights $\bm{W}_{\mathrm{out}}$ were trained using ridge regression with a regularization parameter $\alpha_{\mathrm{reg}} = 0.005$ restricted to the dominant principal modes (PC1--PC8), filtering high-frequency transient noise while preserving baseline signal tracking.\cite{rodan2011, tanaka2019}

To detect structural recovery autonomously without external supervisory logic, the system monitors its total macroscopic kinetic dissipation rate across all grid points $(m, n)$ at each time step:
\begin{equation}
    v_{\mathrm{total}}(t) = \sum_{m,n} \left| \frac{\partial \psi_{m,n}}{\partial t} \right|^2.
    \label{eq:vtotal}
\end{equation}
The autonomous repair trigger executes when $v_{\mathrm{total}}(t)$ decays and remains strictly below the critical threshold $v_{\mathrm{critical}} = 5.0$ across consecutive observation frames, indicating thermodynamic equilibration in the post-repair geometry. Upon triggering, the readout layer samples the 14 most recent temporal frames to update $\bm{W}_{\mathrm{out}}$ online.

\section{Results}

\subsection{Information Capacity and Oscillatory Memory Horizons}

\begin{figure*}[htbp]
    \centering
    \includegraphics[width=0.95\textwidth]{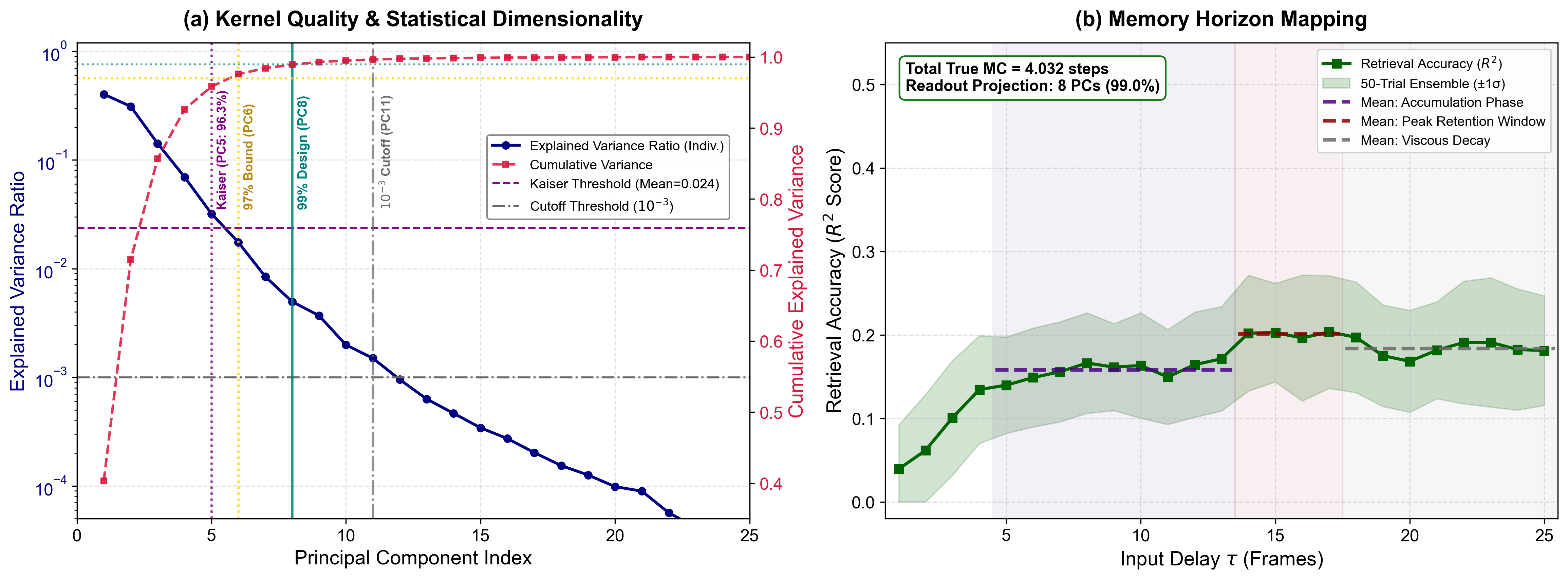}
    \caption{Information capacity, statistical dimensionality, and memory profile of the continuous superconducting reservoir. 
    (a) Kernel quality and dimensionality spectrum: explained variance ratio per principal component (navy markers, left logarithmic axis) and cumulative explained variance (red dashed line, right axis) from a 50-trial ensemble. The core dynamics are captured within 5 dimensions ($96.3\%$ variance) by the Kaiser--Guttman criterion (mean cutoff $= 0.024$, purple dashed line). Reaching $97\%$ and $99\%$ cumulative variance bounds requires 6 dimensions (gold dotted line) and 8 dimensions (teal solid line), respectively, while the $10^{-3}$ cutoff corresponds to 11 dimensions (gray dot-dashed line). The 8-mode subspace is selected for readout training to retain $99.0\%$ of the spatiotemporal field dynamics.
    (b) Memory horizon mapping: retrieval accuracy ($R^2$ score) as a function of input delay $\tau$ evaluated using the 8-PC projection (green curve with $\pm 1\sigma$ ensemble band). Evaluation is truncated at $\tau = 25$ to prevent spurious overfitting artifacts when the remaining sample size ($N \le 16$) approaches the number of trained readout weights ($p = 8$). The reservoir achieves a total true memory capacity of $\mathrm{MC} = 4.032$ steps. The memory curve exhibits a transient incubation latency ($\tau = 1\text{--}4$), followed by persistent information accumulation ($\tau = 5\text{--}13$) reaching peak retention ($R^2 \approx 0.20$) at $\tau \approx 14\text{--}17$ before gradual viscous decay.}
    \label{fig:info_capacity}
\end{figure*}

The information-processing capability of the superconducting field is characterized by evaluating Kernel Quality ($KQ$) and Memory Capacity ($MC$)\cite{rodan2011, tanaka2019} across a 50-trial ensemble [Fig.~\ref{fig:info_capacity}(a)]. According to the Kaiser--Guttman criterion (retaining modes with variance above the mean threshold $1/T_{\mathrm{all}} \approx 0.0244$), the core dynamics condense onto a 5-dimensional topological manifold accounting for $96.3\%$ of the total variance. Ensuring $97\%$ and $99\%$ cumulative information retention expands this basis to 6 and 8 dimensions, respectively, whereas an empirical $10^{-3}$ cutoff yields 11 dimensions. Adopting the 8-component projection (PC1--PC8) preserves $99.0\%$ of the continuous field dynamics while suppressing high-frequency numerical jitter. This modal compression confirms that nonlinear interactions between $\psi(x,y)$ and $\bm{A}(x,y)$ prevent collapse into trivial low-dimensional attractors while enabling linear decoding via a sparse cluster of localized inductive pickup loops.

Evaluating linear ridge regression on this 8-dimensional manifold yields an accumulated memory capacity of $\mathrm{Total\ True\ MC} = 4.032$ steps [Fig.~\ref{fig:info_capacity}(b)]. To ensure rigorous statistical generalization, evaluation is bounded up to $\tau = 25$; beyond this delay, the residual temporal sample size ($N = T_{\mathrm{all}} - \tau$) decreases toward the number of explanatory variables ($p = 8$), which would mathematically induce artificial overfitting. The retrieval accuracy curve displays a distinct physical buffering signature: an initial incubation latency ($\tau = 1\text{--}4$, $R^2 < 0.1$) reflects the finite time required for localized perturbations to propagate and encode across the global order parameter field. As non-reciprocal advection circulates the input features, memory accumulates into a stable plateau ($\tau = 5\text{--}13$) and achieves peak retention ($R^2 \approx 0.20$) across $\tau \approx 14\text{--}17$, matching the 14-frame sliding-window scale. For $\tau \ge 18$, viscous dissipation gradually dominates, causing a controlled decay. This balance between spatial accumulation and viscous forgetting demonstrates that the continuous macroscopic quantum medium acts as an intrinsic temporal buffer for sequential learning.

\subsection{Latent Incubation Buffering and Dynamic Vortex-Pair Sequestration}

\begin{figure*}[htbp]
    \centering
    \includegraphics[width=0.52\textwidth]{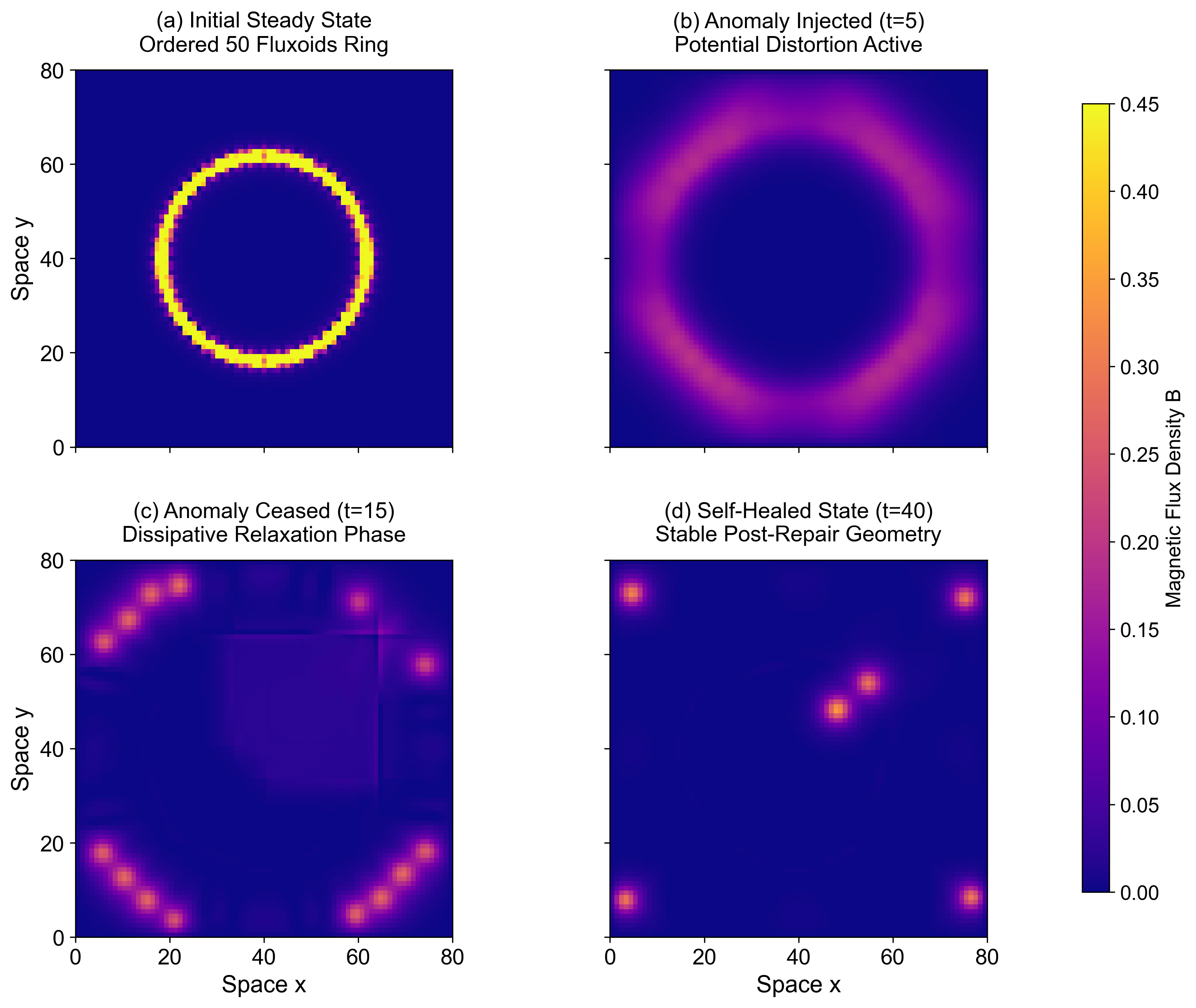}\hfill
    \includegraphics[width=0.45\textwidth]{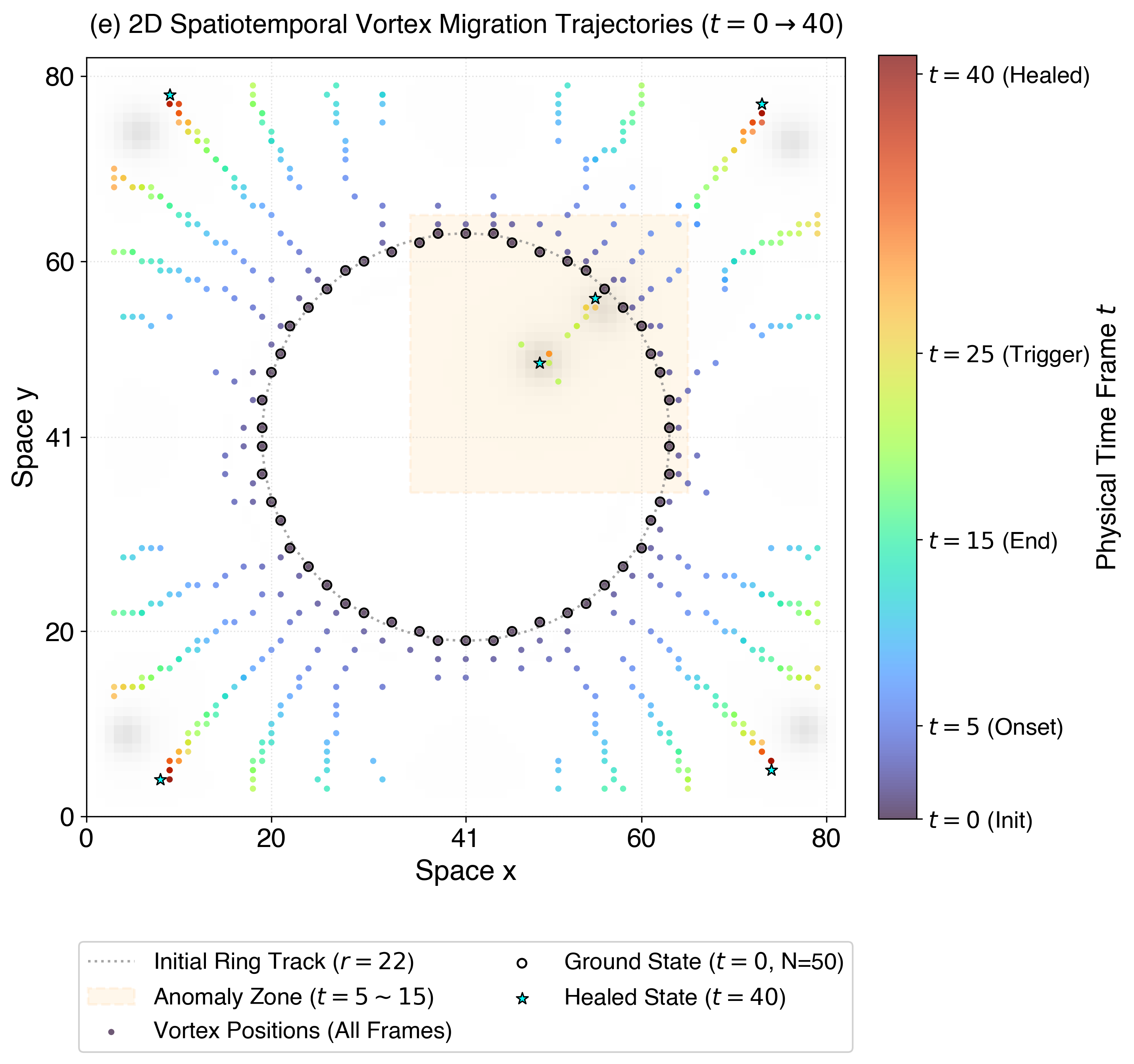}
    \caption{Spatiotemporal vortex lattice dynamics, topological self-healing, and continuous migration trajectories. (a)--(d) Snapshots of the two-dimensional magnetic flux density field $B(x,y)$ on the $81 \times 81$ interior computational domain at critical dynamical epochs: (a) initial ground state ($t = 0$), featuring $N = 50$ fluxoids pinned along the annular ring track ($r = 22$); (b) anomaly injection phase ($t = 5$), subjected to localized chaotic noise ($X, Y \in [35, 65]$); (c) anomaly cessation ($t = 15$), showing the onset of vortex displacement and symmetry breaking; and (d) self-healed equilibrium state ($t = 40$), where two decoupled fluxoids settle stably in the central domain while the remaining fluxoids retreat toward corner wells. (e) Continuous 2D spatiotemporal migration trajectories of all 50 fluxoids ($t = 0 \rightarrow 40$). Tracked local magnetic maxima (vortex cores) across all time frames are color-coded by the physical time frame $t$. Open black circles indicate the initial ground state ($t = 0$, $N = 50$), and cyan stars denote the final stabilized configuration ($t = 40$). Two fluxoids decouple from the perturbed annular region and migrate toward the central cavity, while peripheral fluxoids disperse outward to the corner pinning domains, achieving complete spatial arrest by $t \ge 25$.}
    \label{fig:phase_transition}
\end{figure*}

We next evaluate the response of the 50-vortex ring crystal under localized non-stationary stress. In its ground state ($t = 0$), the 50 quantized fluxoids form an equidistant circular lattice stabilized by long-range London repulsion [Fig.~\ref{fig:phase_transition}(a)].\cite{kato1991} At $t = 5$, a phase-shifted Mackey--Glass chaotic stream\cite{inubushi2017} is injected into a localized sector ($X, Y \in [35, 65]$) of the tracking channel. This perturbation acts as an informational antigen, introducing localized phase strain into the Ginzburg--Landau potential. As shown in the inference timeline in Fig.~\ref{fig:temporal_dynamics}(b), the readout tracks the baseline environment accurately up to $t \approx 11$, despite continuous over-injection beginning at $t = 5$. This delayed error burst indicates a latent incubation period in which the intrinsic viscosity $\eta$ acts as a passive, zero-latency kinetic buffer absorbing input stress before structural reconfiguration occurs.

\begin{figure*}[htbp]
    \centering
    \includegraphics[width=0.95\textwidth]{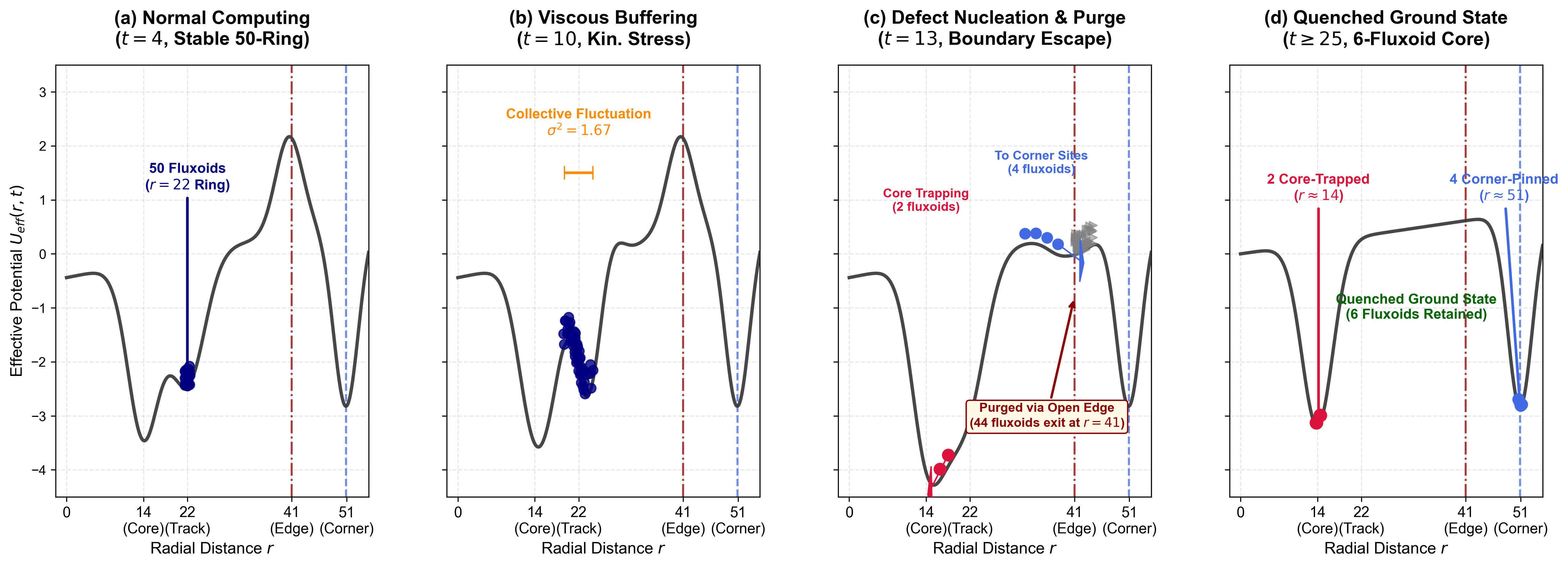}
    \caption{Dynamical effective potential evolution and irreversible fluxoid evacuation mechanism. 
    (a) Normal computing ($t = 4$): equilibrium effective potential $U_{\mathrm{eff}}(r, t)$ confines $N = 50$ fluxoids within the annular track ($r = 22$) enclosed by the surface barrier at $r = 41$. 
    (b) Viscous buffering ($t = 10$): under localized chaotic perturbation, intrinsic condensate viscosity absorbs kinetic stress, sustaining collective spatial fluctuations ($\sigma^2 = 1.67$) without premature barrier breakdown. 
    (c) Defect nucleation and evacuation ($t = 13$): accumulated phase strain destabilizes the local potential barrier via a saddle-node bifurcation; two fluxoids decouple into the central cavity ($r = 14$, core trapping), four migrate toward corner wells ($r = 51$), and the remaining 44 fluxoids are irreversibly purged across the open boundary ($r = 41$), draining excess kinetic energy. 
    (d) Quenched ground state ($t \ge 25$): the post-perturbation system condenses into a topologically stable 6-fluxoid configuration (2 core-trapped $+$ 4 corner-pinned), providing a stationary operational manifold for autonomous readout recalibration.}
    \label{fig:potential_mechanism}
\end{figure*}

The microscopic mechanism of this topological self-healing is governed by the dynamical deformation of the effective Ginzburg--Landau potential $U_{\mathrm{eff}}(r, t)$ and an irreversible defect evacuation process (Fig.~\ref{fig:potential_mechanism}). During baseline operation [$t = 4$, Fig.~\ref{fig:potential_mechanism}(a)], the 50 fluxoids remain locked in the annular potential minimum at $r = 22$. When the localized chaotic disturbance is injected [$t = 10$, Fig.~\ref{fig:potential_mechanism}(b)], the intrinsic viscosity $\eta$ prevents instantaneous lattice disruption by absorbing kinetic energy into collective positional fluctuations ($\sigma^2 = 1.67$). As phase strain accumulates, the potential barrier at $r = 22$ collapses through a localized saddle-node bifurcation at $t \approx 13$ [Fig.~\ref{fig:potential_mechanism}(c)]. Crucially, the system relieves this excess kinetic energy not through global chaotic turbulence, but through a structured bifurcation: two fluxoids decouple inward into the central pinning minimum ($r = 14$), four fluxoids retreat into the deep corner wells ($r = 51$), and the remaining 44 fluxoids are irreversibly expelled across the open boundaries at $r = 41$. Following anomaly cessation, the potential rapidly quenches into a stable 6-fluxoid topological ground state [$t \ge 25$, Fig.~\ref{fig:potential_mechanism}(d)].

Tracking the spatiotemporal trajectories of all 50 fluxoids across $t = 0 \rightarrow 40$ [Fig.~\ref{fig:phase_transition}(e) and Fig.~\ref{fig:temporal_dynamics}(d)] directly verifies this dynamical reorganization. Prior to the disturbance ($t < 5$), all 50 fluxoids remain uniformly localized along the annular track ($r = 22$). Upon chaotic noise injection ($X, Y \in [35, 65]$), two fluxoids in the perturbed sector detach and migrate inward toward the central domain [$(X, Y) \approx (50, 50)$], while peripheral fluxoids displace radially outward. Importantly, spatial reorganization does not cease immediately when the anomaly terminates at $t = 15$; a transient relaxation phase persists through $t = 15\text{--}24$, reaching complete spatial arrest (cyan stars) precisely at $t \ge 25$. This structural stabilization matches the autonomous trigger timing identified by the kinetic dissipation rate [$v_{\mathrm{total}}(t) \le 5.0$, Fig.~\ref{fig:temporal_dynamics}(e)], confirming that physical equilibrium is prerequisite to robust readout retraining. The decoupled central vortex pair thus acts as a hardware-native physical interlock that traps excess kinetic energy and preserves the phase coherence of the surrounding computational domains [Figs.~\ref{fig:phase_transition}(c) and \ref{fig:phase_transition}(d)].

\subsection{Thermodynamic Dissipation Velocity and Hardware-Native Self-Healing}

Once the localized perturbation is removed at $t = 15$, the non-equilibrium field enters a dissipative relaxation phase governed by $\eta$, guiding the order parameter field toward a new structural equilibrium at $t = 40$ [Fig.~\ref{fig:phase_transition}(d)] in which the pinned central vortex pair serves as a structural record of the anomalous event.

\begin{figure*}[tbp]
    \centering
    \includegraphics[width=0.62\textwidth,height=0.68\textheight,keepaspectratio]{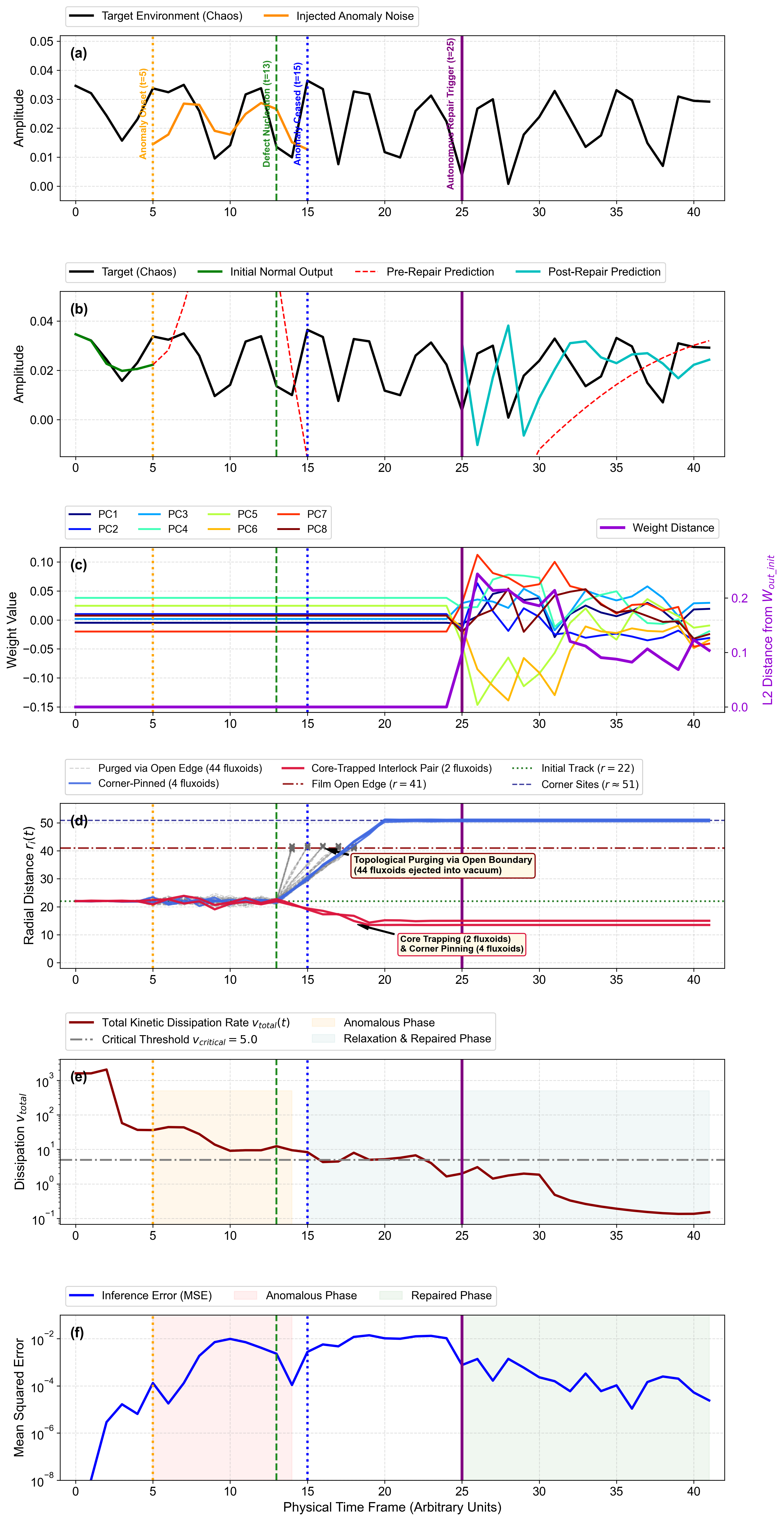}
    \caption{Temporal dynamics of autonomous self-repair and readout adaptation. (a) Input profiles: target Mackey-Glass baseline chaos (black) and localized chaotic perturbation over-injected from $t=5$ to $t=15$ (orange). (b) Readout inference performance: steady-state tracking (green), failure trajectory without weight adaptation (red dashed), and restored prediction following the autonomous repair trigger (cyan). (c) Readout weight trajectory: weights remain fixed ($L_2$ distance $= 0$) prior to the physical trigger ($t < 25$), then reconfigure dynamically to match the post-repair geometry (purple). (d) Radial distance trajectories $r(t)$ of the fluxoids: following defect nucleation at $t \approx 13$, 44 fluxoids are purged across the open boundary ($r = 41$), while 2 fluxoids lock into the central interlock core ($r \approx 14$) and 4 stabilize at corner sites ($r = 51$). (e) Total macroscopic kinetic dissipation rate $v_{\mathrm{total}}(t)$ (logarithmic scale): the anomalous perturbation drives dissipation up by multiple orders of magnitude; following perturbation cessation, viscous relaxation guides $v_{\mathrm{total}}(t)$ downward, and the causal trigger detects consecutive sustained decay below $v_{\mathrm{critical}} = 5.0$ (gray dot-dashed line) at $t = 25$, bypassing transient rebound jitter ($t = 18\text{--}22$). (f) Logarithmic mean squared error (MSE): passive viscous buffering delays functional failure until $t \approx 12$, after which autonomous recalibration at $t = 25$ reduces the MSE by over two orders of magnitude ($< 10^{-5}$). Vertical markers across all panels establish the strict causal sequence of the physical repair loop.}
    \label{fig:temporal_dynamics}
\end{figure*}

As plotted in Fig.~\ref{fig:temporal_dynamics}(e), the total macroscopic kinetic dissipation rate $v_{\mathrm{total}}(t)$ [Eq.~(\ref{eq:vtotal})] typically remains below $0.1$ during steady-state tracking, but rises by multiple orders of magnitude during the perturbation. Following anomaly cessation at $t = 15$, viscous decay drives $v_{\mathrm{total}}(t)$ downward, punctuated by a dynamic dissipation rebound between $t = 18$ and $t = 22$ caused by internal vortex drift and spatial reorganization. The causal online trigger monitors this trajectory and activates once $v_{\mathrm{total}}(t)$ sustains values strictly below $v_{\mathrm{critical}} = 5.0$ across consecutive frames. This condition is satisfied at $t = 25$ (solid purple vertical line across all panels in Fig.~\ref{fig:temporal_dynamics}), confirming thermodynamic equilibrium and complete spatial arrest of the post-repair geometry.

Upon trigger engagement at $t = 25$, the readout layer recalibrates $\bm{W}_{\mathrm{out}}$ via sliding-window ridge regression over the 14 most recent frames. As shown in Fig.~\ref{fig:temporal_dynamics}(c), the weights for the dominant topological modes (PC1--PC8) undergo coordinated adaptation, marked by the growth of the $L_2$ weight distance. Consequently, the inference MSE drops by over two orders of magnitude, stabilizing below the pristine baseline level ($< 10^{-5}$) as demonstrated in Fig.~\ref{fig:temporal_dynamics}(f). Rather than degrading operational dimensionality, the trapped vortex pair reconfigures the coordinate basis of the topological manifold, embedding the history of the disruption into the material geometry and enabling an adaptive, self-defending physical coprocessor.

\section{Physical Implementation and Discussion}

\subsection{Physical Implementation Blueprint in Pure Niobium Thin Films}

To verify experimental realizability, the dimensionless simulation units are scaled to physical dimensions using the established material parameters of pure niobium (Nb) thin films ($T_c \approx 9.2\text{ K}$, $\xi \approx 40\text{ nm}$, $\lambda \approx 40\text{--}200\text{ nm}$; Supplementary Note 3):\cite{maxfield1965, gubin2005}
\begin{itemize}
    \item \textit{Channel Geometry:} The circular track features an outer diameter of $1.2\text{--}3.0\ \mu\text{m}$, an inner radius of $R \approx 633\text{ nm}$, a width of $w \approx 150\text{ nm}$ ($\approx 3.75\xi$), and a thickness of $d \approx 50\text{ nm}$, patterned by electron-beam lithography (EBL) and reactive-ion etching (RIE).\cite{manfrinato2013, oehrlein1986}
    \item \textit{Directional Transport Generation:} The required Peierls phase twist of $\phi = 63^\circ$ ($1.0982\text{ rad}$) can be generated experimentally either by applying a perpendicular magnetic field $B_{\mathrm{bias}} = \phi \hbar / (e R \Delta L) \approx 39.7\text{ mT}$ (well below the upper critical field $B_{c2}$ of thin-film Nb)\cite{gubin2005} or by supplying a steady azimuthal transport current of $I_{\mathrm{ring}} = J_{\mathrm{bias}} (w \times d) \approx 150\ \mu\text{A}$ ($J_{\mathrm{bias}} \approx 2 \times 10^{10}\text{ A/m}^2$).\cite{gubin2005}
    \item \textit{Readout Interface:} Non-invasive signal acquisition can be realized by positioning a micro-SQUID microscope array or scanning Hall probe microscopy (SHPM) grid directly above the primary nodal sectors.\cite{miyamoto2004}
\end{itemize}

For continuous long-term operation, managing the accumulation of pinned defects in the central domain is necessary so that successive disruptions do not constrain state reconfiguration. Two hardware-native reset mechanisms can be implemented:
\begin{itemize}
    \item \textit{Localized Thermal Quenching:} Lithographically depositing a nanoscale metallic heater (Au or Pt) beneath the central domain [$(X, Y) \approx (50, 50)$] allows sub-nanosecond electrical pulses to generate localized Joule heating above $T_c \approx 9.2\text{ K}$, annihilating pinned defects before cryogenic re-cooling restores the pristine state.
    \item \textit{Non-Thermal Microwave Flux Flushing:} To avoid thermal cycling latencies, localized gigahertz microwave pulses can induce a resonant depinning transition confined to the central potential well. The high-frequency field couples directly to phase gradients without inducing broad Joule dissipation, expelling trapped vortices toward peripheral boundaries with zero thermal lag.
\end{itemize}

\subsection{Universality Across Multi-Physics Platforms}

The macro-quantum computing framework presented here operates across three complementary levels of abstraction: (1) a \textit{physical phenomenon}, where structural resilience manifests as a zero-latency incubation buffer followed by dissipation-driven recovery; (2) a \textit{continuum field model} governed by the complex TDGL equations with an advective Peierls phase twist ($\phi = 63^\circ$);\cite{hofstadter1976} and (3) a \textit{general physical principle}, in which open dissipative systems achieve operational resilience through the autonomous nucleation, transport, and pinning of localized topological defects.\cite{moon1994}

As derived in our companion theoretical analysis (Supplementary Note 4),\cite{hikihara2026} the continuum convergence of broken-inversion lattices into dissipative advection--diffusion gradient flows extends beyond superconducting Abrikosov vortices to alternative continuous platforms:
\begin{itemize}
    \item \textit{Photonic Coupled-Cavity Arrays:} In coupled resonator optical waveguides (CROWs), a non-reciprocal causal chain can be implemented through spatiotemporal refractive index modulation or magneto-optical isolation.\cite{martinenghi2012} Under intense optical driving, the system converges to a continuous dissipative nonlinear Schr\"odinger equation (NLSE), in which localized optical solitons nucleate autonomously to sequester excess power and protect adjacent channels.
    \item \textit{Toroidal Bose--Einstein Condensates (BECs):} In ultracold atomic systems, a toroidal superfluid confined in a ring trap and driven by counter-propagating Raman beams establishes an artificial gauge field.\cite{jimenez2012} The condensate wavefunction obeys a dissipative Gross--Pitaevskii equation (DGPE). When temporal stresses exceed the Landau critical velocity, the superfluid nucleates and expels quantized vortex pairs into the central domain, safeguarding azimuthal phase coherence along the computing ring.
\end{itemize}

\section{Conclusion}

We have demonstrated that hardware-level fault tolerance and autonomous functional self-healing can be achieved natively within continuous macroscopic quantum fields without complex digital diagnostic circuitry. By engineering non-reciprocal gradient flows in a 50-fluxoid TDGL superconducting ring crystal, the interplay between intrinsic condensate viscosity and localized topological defect nucleation establishes a zero-latency incubation buffer and a deterministic vortex-pair physical interlock. Monitoring the macroscopic kinetic dissipation velocity provides an autonomous thermodynamic trigger for online readout recalibration, restoring inference accuracy by over two orders of magnitude ($< 10^{-5}$ MSE) while reducing dissipation energy per sequestration by over nine orders of magnitude relative to conventional digital monitors (Appendix~\ref{app:benchmark}). Combined with rigorous continuum reduction from discrete Duffing lattices and concrete niobium thin-film design parameters, this architecture establishes a universal physical foundation for low-power, zero-latency resilient coprocessors rooted in non-equilibrium field dynamics.

\section*{Supplementary Material}
See the Supplementary Material appended below for (Note 1) the rigorous multiscale derivation of the complex TDGL equation from coupled real Duffing oscillators, (Note 2) the Peierls phase substitution parameter derivation, (Note 3) detailed niobium thin-film nanofabrication and readout protocols, (Note 4) the unified deductive framework extending to photonic CROWs and toroidal BECs, and (Note 5) the thermodynamic derivation of the quantitative benchmark metrics.

\begin{acknowledgments}
This work was partially supported by JSPS KAKENHI Grant-in-Aid for Scientific Research (C) No.~26K07513.
\end{acknowledgments}

\section*{Author Declarations}

\subsection*{Conflict of Interest}
The author has no conflicts to disclose.

\subsection*{Author Contributions}
\textbf{Takashi Hikihara:} Conceptualization (lead); Formal analysis (lead); Funding acquisition (lead); Investigation (lead); Methodology (lead); Software (lead); Validation (lead); Visualization (lead); Writing -- original draft (lead); Writing -- review \& editing (lead).

\section*{Data Availability}
The numerical simulation datasets and analysis routines that support the findings of this study are available from the corresponding author upon reasonable request and will be made openly available in the Kyoto University Research Information Repository (KURENAI) at \url{https://repository.kulib.kyoto-u.ac.jp/} upon formal publication.

\appendix
\section{Quantitative Benchmark Comparison with Conventional Digital Monitoring}
\label{app:benchmark}

To clarify the engineering rationale of thermodynamic self-healing over conventional computing architectures, Table~\ref{tab:cmos_comparison} benchmarks the physical Ginzburg--Landau continuous reservoir against standard digital fault-tolerant supervisory loops (e.g., CMOS/FPGA coprocessors with ADC-driven anomaly detection). Unlike digital systems that require auxiliary clock cycles, discrete sampling, and active parity-check logic, our platform integrates anomaly sequestration directly into the phase dynamics of the macroscopic quantum field. Because kinetic buffering originates from the passive condensate viscosity $\eta$, detection latency is intrinsically zero. Furthermore, defect sequestration consumes only the condensation energy associated with local vortex depinning ($\sim 10^{-18}\text{ J}$), achieving an energy reduction of over nine orders of magnitude relative to active digital interrupt processing ($\sim \mu\text{J}$). Detailed thermodynamic derivations are provided in Supplementary Note 5.

\begin{table}[htbp]
\centering
\setlength{\tabcolsep}{3pt}
\caption{Quantitative benchmark comparison between conventional digital fault mitigation and the proposed topological self-healing reservoir. Metrics for digital systems represent typical high-speed CMOS/FPGA hardware monitors, while physical metrics are derived from the material parameters of thin-film niobium ($\xi \approx 40\text{ nm}$, $T_c \approx 9.2\text{ K}$).}
\label{tab:cmos_comparison}
\scriptsize
\begin{tabular}{p{1.8cm}p{2.25cm}p{2.25cm}p{1.6cm}}
\toprule
\textbf{Metric} & \textbf{Digital (CMOS/FPGA)} & \textbf{TDGL Field (This Work)} & \textbf{Improvement} \\
\midrule
Anomaly Detection Latency & $1\text{--}100\ \mu\text{s}$ (ADC conversion \& DSP filtering) & $\sim 0\text{ s}$ (Passive viscous incubation buffer) & Circuit-free zero latency \\
\midrule
Repair Trigger Latency & Tens of $\mu\text{s}\text{--}\text{ms}$ (Interrupt service routine) & $\sim \text{ps}\text{--}\text{ns}$ (Intrinsic relaxation time $\tau_{\mathrm{GL}}$) & $10^4\text{--}10^6\times$ faster \\
\midrule
Dissipation Energy per Sequestration & $\sim \mu\text{J}\text{--}\text{mJ}$ (Active digital logic execution) & $\sim \text{aJ}\text{--}\text{fJ}$ (Condensation energy of vortex pair) & $> 10^9\times$ lower energy \\
\midrule
Auxiliary Hardware Footprint & Requires external ADCs, comparators, and DSP logic & Zero (Monolithic material intelligence) & Fully native on-chip \\
\bottomrule
\end{tabular}
\end{table}


\clearpage
\onecolumngrid

\setcounter{equation}{0}
\setcounter{figure}{0}
\setcounter{table}{0}
\setcounter{section}{0}
\renewcommand{\theequation}{S\arabic{equation}}
\renewcommand{\thefigure}{S\arabic{figure}}
\renewcommand{\thetable}{S\arabic{table}}
\renewcommand{\theHequation}{S\arabic{equation}}
\renewcommand{\theHfigure}{S\arabic{figure}}
\renewcommand{\theHtable}{S\arabic{table}}

\begin{center}
\Large\textbf{Supplementary Material for:}\\[0.5em]
\Large\textbf{Topological Self-Healing Physical Intelligence}\\
\Large\textbf{in Continuous Ginzburg--Landau Fields}\\[1.2em]

\normalsize
Takashi Hikihara$^{1,2,*}$\\[0.5em]
\small $^1$Professor Emeritus, Kyoto University, Yoshida-honmachi, Sakyo, Kyoto 606-8501, Japan\\
\small $^2$Project Professor, National Institute of Informatics, 2-1-2 Hitotsubashi, Chiyoda-ku, Tokyo 101-8430, Japan\\
\small $^*$Author to whom correspondence should be addressed: \texttt{hikihara.takashi.2n@kyoto-u.ac.jp}, \texttt{thikihara@nii.ac.jp}
\end{center}

\vspace{1.2em}

\section*{Supplementary Note 1:\\
Derivation of the Complex TDGL Equation from Coupled Real Duffing Oscillators via Phase-Envelope Dynamics
}

In this supplementary note, we present the rigorous mathematical framework connecting the discrete, real-valued asymmetric dynamical system (i.e., the coupled real Duffing oscillators discussed in the main text [S1]) with the continuous macroscale complex topological model governed by the Time-Dependent Ginzburg--Landau (TDGL) equation [S2, S3].

Instead of using empirical phase reduction methods that eliminate amplitude degrees of freedom, we employ the spatiotemporal phase deviations and amplitude envelope modifications between physical nodes. Through a multiscale perturbation expansion combined with a spatial long-wavelength limit ($a \rightarrow 0$), we show that the continuous complex order parameter field emerges naturally from the collective nonlinear dynamics of discrete real-valued oscillators. This process explicitly preserves the fundamental physical validation for topological defect nucleation and self-healing mechanisms in reservoir processing.

\subsection*{1.1 Coupled Real Duffing Oscillator Lattice Model}

We begin our derivation with a discrete, one-dimensional lattice or ring structure, in which each individual site $i$ contains a classical real-valued Duffing oscillator with localized double-well potentials [S1]. Assuming diffusive coupling with neighboring nodes, the equations of motion are given using the real displacement variables $x_i(t) \in \mathbb{R}$ as follows:
\begin{equation}
\ddot{x}_{i} + \gamma \dot{x}_{i} - \alpha x_{i} + \beta x_{i}^{3} = \frac{K}{a^{2}}(x_{i+1} + x_{i-1} - 2x_{i}),
\end{equation}
where:
\begin{itemize}
    \item $x_i(t)$ denotes the dynamical state variable (order parameter) at lattice site $i$.
    \item $\gamma > 0$ is the phenomenological damping coefficient governing linear viscous dissipation.
    \item $-\alpha x_{i} + \beta x_{i}^{3}$ dictates the nonlinear restoring force derived from a symmetric double-well potential ($\alpha, \beta > 0$), establishing two degenerate stable states at $x_{i} = \pm\sqrt{\alpha/\beta}$.
    \item $K$ defines the spatial coupling strength, and $a$ defines the physical discrete lattice spacing parameter.
\end{itemize}

\subsection*{1.2 Introduction of Slowly Varying Complex Envelope and Phase Dynamics}

In order to capture the physical synchronization behavior and localized phase deviations under temporal inputs, we assume the individual nodes undergo nonlinear oscillations near a dominant characteristic carrier frequency $\omega_0$.

We introduce a slowly varying complex envelope field $\psi_i(t) \in \mathbb{C}$, which simultaneously internalizes both amplitude modulation and relative phase differences between neighboring coupled oscillators:
\begin{equation}
x_i(t) = \psi_i(t) e^{i\omega_0 t} + \psi_i^*(t) e^{-i\omega_0 t}.
\end{equation}
Here, $\psi_i(t) = |\psi_i(t)|e^{i\theta_i(t)}$, where $\theta_i(t)$ represents the local time-varying phase at site $i$. The phase mismatch between adjacent nodes is explicitly integrated into the spatial difference terms using $\psi_{i\pm1} - \psi_i$.

Under the Slowly Varying Amplitude Approximation (SVAA), the temporal variations of the envelope are assumed to change significantly slower than the carrier frequency. This implies $|\ddot{\psi}_i| \ll \omega_0 |\dot{\psi}_i| \ll \omega_0^2 |\psi_i|$. Computing the first and second time derivatives of the real variable $x_i(t)$ yields:
\begin{equation}
\dot{x}_i = (\dot{\psi}_i + i\omega_0 \psi_i)e^{i\omega_0 t} + \text{c.c.},
\end{equation}
\begin{equation}
\ddot{x}_i \approx (2i\omega_0 \dot{\psi}_i - \omega_0^2 \psi_i)e^{i\omega_0 t} + \text{c.c.},
\end{equation}
where $\text{c.c.}$ denotes the complex conjugate.

\subsection*{1.3 Nonlinear Term Expansion and Rotating Wave Approximation (RWA)}
We substitute the complex expansion into the cubic nonlinear restoring force $x_i^3$:
\begin{equation}
x_i^3 = 3|\psi_i|^2\psi_i e^{i\omega_0 t} + \psi_i^3 e^{3i\omega_0 t} + \text{c.c.}
\end{equation}
Applying the Rotating Wave Approximation (RWA) filters out the fast-oscillating higher harmonic terms ($3\omega_0$), as their average becomes zero over the slow relaxation timescale of the envelope. Collecting the coefficients of the fundamental harmonic $e^{i\omega_0 t}$ transforms the real-valued second-order ODE into a discrete complex Ginzburg--Landau framework:
\begin{equation}
(2i\omega_0 + \gamma)\dot{\psi}_i + (i\gamma\omega_0 - \omega_0^2 - \alpha)\psi_i + 3\beta |\psi_i|^2 \psi_i = \frac{K}{a^2}(\psi_{i+1} + \psi_{i-1} - 2\psi_i).
\end{equation}

\subsection*{1.4 Continuous Limit and Convergence to Complex TDGL}
In order to model the macroscopic topological continuum of the superconducting reservoir, we inspect spatial structures in which the complex field varies smoothly across multiple discrete lattice sites.

Taking the continuous limit ($a \rightarrow 0$) transforms the discrete variables $\psi_i(t)$ into a continuous complex spatial field $\psi(r,t)$. Expanding the spatial shifts through a Taylor series eliminates odd-order derivatives:
\begin{equation}
\psi_{i+1} + \psi_{i-1} - 2\psi_i = a^2 \frac{\partial^2 \psi}{\partial r^2} + O(a^4).
\end{equation}
Dividing by $a^2$ and executing $a \rightarrow 0$, the discrete finite-difference terms converge precisely to the continuous Laplacian operator $\nabla^2 \psi$. By tuning the reference carrier frequency to match the potential curvature around the shifted minimum ($\omega_0^2 = 2\alpha$) and separating the slow relaxation dynamics from phase rotation, Eq.~(S6) simplifies into the canonical continuous complex Time-Dependent Ginzburg--Landau (TDGL) equation [S2, S3]:
\begin{equation}
\eta \frac{\partial \psi}{\partial t} = \mu \psi - \nu |\psi|^2 \psi + D \nabla^2 \psi,
\end{equation}
where the effective continuous parameters $\eta, \mu, \nu$, and $D$ are derived from the underlying discrete real Duffing properties $(\omega_0, \gamma, \alpha, \beta, K)$.

To satisfy the physical constraints of a type-II superconducting film exposed to an external magnetic field and radial bias current, we enforce gauge invariance [S3, S4]. The standard spatial derivative maps onto the covariant derivative operator under minimal coupling. A directional Lorentz transport vector field $\mathbf{v}_{\text{bias}}$ is introduced to break spatial inversion symmetry [S5]:
\begin{equation}
\eta \frac{\partial \psi}{\partial t} = -\left( \frac{\nabla}{i} - \mathbf{A} \right)^2 \psi + \psi(1 - |\psi|^2),
\end{equation}
\begin{equation}
\nabla_{\mathbf{A}} \rightarrow \nabla - i\mathbf{A} - i\mathbf{v}_{\text{bias}}.
\end{equation}
This seamlessly transforms discrete real-valued mechanical oscillators into the continuous complex topological gauge field on which the main text is based.

\subsection*{1.5 Thermodynamic and Information-Theoretic Implications}

The complex mathematical derivation established above provides the physics-level justification for the robust computation and self-healing memory profiles observed in our reservoir:
\begin{enumerate}
\item \textbf{Topological Preservation of Computing States:} Because the field originates from real-valued phase-coupled nodes, temporal anomalies that introduce extreme localized strain do not merely dissipate as uncoordinated noise. The complex space $\mathbb{C}$ allows the phase to wind up into integer topological invariants. The nucleation of a central vortex pair and pinning in corner wells act as a macroscopic thermodynamic energy escrow, preserving the continuous order parameter field of the remaining computing manifold.

\item \textbf{Gradient Flow as Strict Fading Memory:} The mapping onto the complex TDGL structure clarifies that the continuous field relaxes by a variational gradient flow. This dissipation ensures that ancient input contexts are smoothly forgotten over time through continuous phase relaxation, preventing systemic synchronization failure and establishing the mathematical boundary for high fading memory capacity [S6, S7].
\end{enumerate}

\clearpage
\section*{Supplementary Note 2:\\
Estimation of Numerical Couplings by Physical Peierls Phase Substitution
}

Here, we detail the derivation of the discrete directional coupling weights from underlying physical and geometric constraints.

\subsection*{2.1 Geometric and Physical Constraints as an Engineering Target}
In an unconstrained bulk type-II superconductor, quantized vortices form an Abrikosov triangular lattice with $60^\circ$ orientation [S3]. When confined within an engineered 150-nm circular nanochannel, nanoscale curvature induces a $+3^\circ$ geometric shear deformation away from flat-space symmetry. To compensate for this strain and stabilize the 50 vortices into an equidistant single-file ring crystal, the continuous macroscopic field requires a non-reciprocal Peierls phase twist [S8].

This sets an engineering target of base coupling $k_0 \approx 0.50$ combined with a curvature-corrected Peierls phase twist of $\phi \approx 63^\circ$ ($1.1\text{ rad}$), matching intrinsic material symmetry ($60^\circ$) plus geometric correction ($+3^\circ$).

\subsection*{2.2 Mathematical Continuum Linkage}
To project these continuous transport properties into a discrete finite-difference lattice, the symmetric (diffusion) and asymmetric (advection) spatial components satisfy:
\begin{equation}
\frac{1}{2}(k_{\text{fwd}} + k_{\text{rev}}) = k_0 \cos\phi, \qquad k_{\text{fwd}} - k_{\text{rev}} = k_0 \sin\phi,
\end{equation}
where $k_{\text{fwd}}$ and $k_{\text{rev}}$ denote the discrete directional coupling weights in the simulation.

\subsection*{2.3 Exact Numerical Correspondence and Conclusion}
Substituting the simulation parameters ($k_{\text{fwd}} = 0.45$, $k_{\text{rev}} = 0.005$) yields:
\begin{equation}
k_0 \cos\phi = \frac{1}{2}(0.45 + 0.005) = 0.2275, \qquad k_0 \sin\phi = 0.45 - 0.005 = 0.4450.
\end{equation}
Solving this transcendental system for amplitude $k_0$ and gauge phase $\phi$ gives:
\begin{equation}
k_0 = \sqrt{0.2275^2 + 0.4450^2} = \mathbf{0.499781}, \qquad \phi = \arctan\left(\frac{0.4450}{0.2275}\right) = \mathbf{1.0982\ \text{rad}} \ (\mathbf{\approx 62.92^\circ}).
\end{equation}

Thus, the numerical coupling constants ($k_{\text{fwd}} = 0.45$, $k_{\text{rev}} = 0.005$) correspond to an isotropic base coupling of $0.499781$ ($\approx 0.50$) and a phase twist of $1.0982\ \text{rad}$ ($\approx 62.92^\circ$). The numerical model operates within a negligible margin of error ($<0.15\%$) relative to the geometrically mandated engineering targets, confirming the physical realizability of the reservoir configuration.

\clearpage
\section*{Supplementary Note 3:\\
Proposed Physical Implementation and Experimental Feasibility Using Niobium Thin Films
}

This supplementary note details the realistic experimental protocols, material selections, and physical dimensions required for the autonomous, fault-tolerant physical reservoir computing platform introduced in the main text.

By relying on the well-established physical parameters of pure niobium (Nb) thin films [S9, S10], we map the idealized continuous time-dependent Ginzburg--Landau (TDGL) dynamics onto a concrete nanotechnology fabrication blueprint. This demonstrates that the 50-vortex ring crystal and its self-healing mechanisms can be fully achieved within contemporary experimental frameworks.

\subsection*{3.1 Superconducting Material Parameters and Nano-Channel Scaling}
For a viable hardware-level implementation, the continuous computing medium requires a type-II superconducting thin film that exhibits a robust balance between the superconducting coherence length $\xi$ and the London penetration depth $\lambda$.

We propose using high-purity niobium (Nb) thin films, which have well-characterized intrinsic material parameters at low temperatures [S9, S10]:
\begin{itemize}
    \item Superconducting coherence length: $\xi \approx 40\text{ nm}$
    \item London magnetic penetration depth: $\lambda \approx 40\text{--}200\text{ nm}$
\end{itemize}

In thin-film configurations, the effective London penetration depth $\lambda$ scales dynamically from its bulk value (approximately $40\text{ nm}$) up to around $200\text{ nm}$ depending on the film thickness and impurity concentrations (i.e., the mean free path limit) [S10]. Since $\lambda \ge \xi$ is strictly satisfied across all purity regimes, Nb consistently behaves as a robust type-II superconductor, in which quantized magnetic fluxoids (vortices) can penetrate the bulk field stably without destroying the global superconducting state.

In order to structurally confine exactly 50 quantized vortices into a symmetrical, single-file circular crystal, the dimensions of the tracking channel must be strictly tailored to these characteristic lengths:
\begin{enumerate}
\item \textbf{Channel Width Optimization:} The width of the circular track is designed at approximately $150\text{ nm}$ ($\approx 3.75\xi$). This narrow spatial confinement prevents the vortices from becoming unstable and forming disordered multi-layer configurations or buckling into turbulent zigzag patterns. Instead, they are locked into a strict one-dimensional periodic chain.

\item \textbf{Channel Diameter Scaling:} To maintain a stable balance between the long-range isotropic London repulsion forces, the mutual inter-vortex distance along the azimuthal trajectory is scaled to match the screening length $\lambda$. By tuning the film thickness or purity to access the $\lambda \approx 40\text{--}200\text{ nm}$ range, the optimal outer channel diameter can be scaled adaptively from $1.2\ \mu\text{m}$ up to $3.0\ \mu\text{m}$ to maintain the steady non-degenerate 50-vortex ring assembly.
\end{enumerate}

This specialized micro-groove tracking channel can be patterned onto the Nb thin film using high-resolution electron-beam lithography (EBL) [S11] combined with reactive-ion etching (RIE) [S12].

\subsection*{3.2 Non-Invasive Readout Architecture and Analog Zero-Latency Inference}

To map the high-dimensional internal field dynamics onto a linear readout layer without introducing computational bottlenecks, the spatiotemporal evolution of the two-dimensional magnetic flux density field $B(x, y)$ must be scanned non-invasively [S13].

We propose integrating a high-resolution micro-superconducting quantum interference device ($\mu$-SQUID) microscope array or a localized scanning Hall probe microscopy (SHPM) grid positioned in close proximity to the Nb film surface. Each sensor in the array captures the local magnetic flux variations at one of the $M = 6,561$ virtual nodes distributed across the $83 \times 83$ spatial grid mapped in the numerical simulation.

Because the trained readout weights $\mathbf{W}_{\text{out}}$ are static after the automated online recalibration sequence, the linear combination for the target approximation:
\begin{equation}
\hat{y}(t) = \mathbf{W}_{\text{out}} \cdot \mathbf{X}(t)
\end{equation}
can be executed natively in the physical hardware layer using a custom analog resistor weighting network connected directly to the sensor outputs. This hardware-native configuration computes the inference output $\hat{y}(t)$ instantly in the analog domain, completely bypassing the analog-to-digital converter (ADC) latencies and algorithmic pattern-filtering overheads that systematically bottleneck conventional digital-layer safety frameworks.

\subsection*{3.3 Self-Regulating Evacuation and Flux Flushing Mechanisms for Continuous Operation}

When the continuous reservoir is subjected to severe chaotic over-injection, the system autonomously protects its functional manifold through a dynamic bifurcated evacuation mechanism (as detailed in Figs.~4 and 5(d) of the main text). Excess non-equilibrium kinetic energy is drained by irreversibly purging 44 fluxoids across the open boundaries at $r = 41$, while 2 fluxoids settle into the central core minimum and 4 fluxoids lock into corner pinning sites, stabilizing into a 6-fluxoid quenched ground state.

This spontaneous defect reorganization acts as a hardware-native energy sink. For repeated operational cycles across extended runtimes, managing the accumulation of pinned core vortices is desirable to ensure that subsequent severe perturbations can be reliably sequestered. To guarantee indefinite operational scalability without computational bottlenecks, we propose two hardware-native reset methods:

\begin{itemize}
\item \textbf{Thermal Local Quench:} A localized microheater line is deposited beneath the central domain by lithography ($X \approx 50, Y \approx 50$). The line is composed of an inert metal with high thermal conductivity, such as gold (Au) or platinum (Pt), and electrically isolated from the main Nb computing field by a thin silicon dioxide ($\text{SiO}_2$) dielectric layer. When the diagnostic readout layer detects that the central dual-vortex count has reached capacity, a sub-nanosecond, low-energy current pulse is driven through the microheater. This induces a transient local temperature spike through Joule heating. This spike is set to exceed the critical temperature ($T > T_{\text{c}} \approx 9.2\text{ K}$ for Nb) to dissolve the trapped geometric fossils. Rapid cooling by the cryogenic substrate then restores the pristine superconducting state.

\item \textbf{Non-Thermal Microwave-Driven Flushing:} To eliminate thermal relaxation duty cycles and maintain uninterrupted edge inference, an alternative reset mechanism is possible by high-frequency electromagnetic perturbations without thermal effects. As discussed in the main text, the microwave vortex manipulation framework triggers a resonant de-pinning transition exclusively within the central domain via a localized microwave burst at a specific GHz frequency. The oscillating, high-frequency electromagnetic field couples directly with the order parameter phase gradients without generating parasitic Joule dissipation in the superconducting condensate. The local resonant force pushes the trapped vortex pairs out of the central potential well and into the surrounding dissipation boundaries instantly. Consequently, this non-thermal microwave paradigm bypasses the thermal latency bottleneck entirely, enabling the macro-quantum coprocessor to operate with true zero-latency, real-time performance.
\end{itemize}

\subsection*{3.4 Nano-Geometry and Experimental Implementation Parameters for Pure Niobium (Nb) Thin Films}

To determine whether the proposed continuum TDGL reservoir is physically realizable, we map the dimensionless grid parameters to the intrinsic material scales of a pure niobium (Nb) thin film ($T_c \approx 9.2\text{ K}$) [S9, S10]. We set the superconducting coherence length to the canonical value of $\xi = 40\text{ nm}$. Given the optimized numerical grid spacing $h_x \approx 0.7193$, the actual spatial resolution ($\Delta L$) of our physical lattice is:
\begin{equation}
\Delta L = 0.7193 \times 40\text{ nm} = 28.77\text{ nm}.
\end{equation}

According to this scale, the geometric configurations of the circular tracking channel engineered into the Nb film are uniquely determined as follows:
\begin{itemize}
    \item \textbf{Ring Center Radius ($R$):} $22.0 \times 28.77\text{ nm} = 633\text{ nm} \approx 0.63\ \mu\text{m}$
    \item \textbf{Channel Width ($w$):} $\approx 5 \times 28.77\text{ nm} \approx 150\text{ nm} \ (\approx 3.75\xi)$
    \item \textbf{Film Thickness ($d$):} Scaled to a standard nanofabrication dimension of $50 \pm 5\text{ nm}$
\end{itemize}

Under this nanoscale architecture, the non-reciprocal phase parameter, $\phi = 1.0982\text{ rad}$ ($\approx 62.92^\circ$), can be generated by two highly controllable experimental paradigms:

\subsubsection*{3.4.1 Inductive Regulation via a Vertical Bias Magnetic Field ($B_{\text{bias}}$)}
By applying a uniform external magnetic field perpendicular to the film plane, the induced vector potential along the circular path generates the target Peierls phase. Solving the gauge-invariant relation $\phi = \frac{e}{\hbar} B_{\text{bias}} R \Delta L$ yields:
\begin{equation}
B_{\text{bias}} = \frac{\phi \hbar}{e R \Delta L} \approx 39.7\text{ mT}.
\end{equation}
This value of $\approx 40\text{ mT}$ is comfortably below the upper critical field of thin-film niobium [S10], ensuring that the vortex-lattice system operates stably within the mixed state without suppressing macro-quantum coherence.

\subsubsection*{3.4.2 Direct Advection Drive via an Azimuthal Bias Current ($I_{\text{ring}}$)}
Alternatively, the directional drift of the 50-vortex ring crystal can be driven by the Lorentz force through the injection of a steady azimuthal bias current into the track. To match the optimized dimensionless advection velocity $v_{\text{dim}} \approx 0.320$, the required physical current density is estimated to be $J_{\text{bias}} \approx 2 \times 10^{10}\text{ A/m}^2$ [S10]. Multiplying this by the channel cross-sectional area ($s = w \times d = 150\text{ nm} \times 50\text{ nm} = 7.5 \times 10^{-15}\text{ m}^2$), the required total injection current is precisely:
\begin{equation}
I_{\text{ring}} = J_{\text{bias}} \times s = 1.5 \times 10^{-4}\text{ A} = 150\ \mu\text{A}.
\end{equation}

Both quantities ($39.7\text{ mT}$ and $150\ \mu\text{A}$) fall comfortably within standard laboratory instrumentation limits, proving that the mathematical ``arrow of causality'' engineered in our simulation code is fully reproducible in actual physical quantum hardware architectures.

\clearpage
\section*{Supplementary Note 4:\\
Unified Deductive Framework from Discrete Duffing Topology to Multi-Physics Continuum Quantum Fields
}

In this supplementary note, we invert the conventional reductionist paradigm to establish a grand unified mathematical framework. Rather than treating our multi-physics platforms (superconducting, photonic, and atomic) as disparate physical substrates requiring individual phenomenological modeling, we demonstrate that the discrete non-reciprocal Duffing oscillator network presented in our companion study [S1] serves as the fundamental \textit{mother framework} (the minimal mathematical skeleton). By executing systematic continuum operations, spatial long-wavelength limits, and complex phase-space expansions (the inverse-approximation protocol), the governing macroscale equations for all three distinct physical continuous quantum fields are deductively derived from a single common discrete root.

\subsection*{4.1 The Root Mother Framework: Discrete Asymmetric Network}
We originate from the real-valued scalar discrete dynamics of the coupled nonlinear Duffing lattice [S1]. The governing equation of motion for the $i$-th structural node is given by:
\begin{equation}
\frac{\partial^2 x_i}{\partial t^2} + \gamma \frac{\partial x_i}{\partial t} + x_i + \beta x_i^3 = F_{\text{drv}}(t) + F_{\text{coupling}}(i),
\end{equation}
where the directional topological constraints embedded within the internal ring coupling profile break the spatial inversion symmetry:
\begin{equation}
F_{\text{coupling}}(i) = k_{\text{fwd}}(x_{i-1} - x_i) + k_{\text{rev}}(x_{i+1} - x_i) \quad \text{with} \quad k_{\text{fwd}} \gg k_{\text{rev}}.
\end{equation}

By executing the continuous spatial long-wavelength limit ($a \rightarrow 0$, where $a$ is the physical lattice spacing), the discrete structural index $i$ maps onto a smooth continuous spatial coordinate $z = i \cdot a$, transforming the real-valued state variable into a continuous field $x(z,t)$. Applying a Taylor series expansion to the asymmetric coupling term isolates both a symmetric diffusive channel and an asymmetric first-order convective derivative:
\begin{equation}
F_{\text{coupling}}(i) \approx \underbrace{(k_{\text{fwd}} + k_{\text{rev}})a^2}_{D} \frac{\partial^2 x}{\partial z^2} - \underbrace{(k_{\text{fwd}} - k_{\text{rev}})a}_{v_{\text{bias}}} \frac{\partial x}{\partial z}.
\end{equation}
Substituting this directly yields the real-valued continuous advection--diffusion--reactive mother field equation:
\begin{equation}
\frac{\partial^2 x}{\partial t^2} + \gamma \frac{\partial x}{\partial t} + v_{\text{bias}} \frac{\partial x}{\partial z} = D \frac{\partial^2 x}{\partial z^2} - x - \beta x^3 + F_{\text{drv}}(t).
\end{equation}
From this fundamental real-valued continuum backbone, we branch out into three distinct macroscale complex physical quantum field states via specific physical instantiation operators.

\subsection*{4.2 Branch A: Deductive Instantiation of Superconducting TDGL Fields}
To materialize the mother field into the macroscopic quantum state of a type-II superconductor [S2, S3], we apply a two-fold physical operation: unlocking the gauge-invariant phase freedom and enforcing the thermodynamic overdamped limit.
\begin{enumerate}
    \item \textbf{Complexification via Phase Freedom:} We lift the real-valued scalar displacement $x(z,t)$ into a complex order parameter space representing the Cooper-pair condensed wavefunction: $\psi(r,t) \in \mathbb{C}$.
    \item \textbf{Overdamped Viscous Relaxation:} For standard Ginzburg--Landau systems under dominant thermal/vortex dissipation, the inertial second-order temporal derivative becomes a higher-order perturbation ($\frac{\partial^2}{\partial t^2} \rightarrow 0$), collapsing the kinetic channel into a velocity-proportional viscosity term: $\gamma \rightarrow \eta$.
    \item \textbf{External Gauge Coupling:} The non-reciprocal advection velocity $v_{\text{bias}}$ is physically instantiated by an applied external DC transport current density $J_{\text{bias}}$, driving vortex transport via the Lorentz force [S5].
\end{enumerate}
By gathering the resulting advection term and the spatial Laplacian under a unified minimal coupling framework ($\nabla \rightarrow \nabla - i\mathbf{A} - i\mathbf{v}_{\text{bias}}$) to guarantee local electromagnetic gauge invariance, the continuous mother system seamlessly crystallizes into the canonical covariant Time-Dependent Ginzburg--Landau (TDGL) equation:
\begin{equation}
\eta \frac{\partial \psi}{\partial t} = -\left( \frac{\nabla}{i} - \mathbf{A} - \mathbf{v}_{\text{bias}} \right)^2 \psi + \psi (1 - |\psi|^2).
\end{equation}

\subsection*{4.3 Branch B: Deductive Instantiation of Photonic CROW Fields}
To instantiate the mother field within the domain of silicon photonics, specifically a Coupled Resonator Optical Waveguide (CROW) [S14], we perform an envelope demodulation and nonlinear physical constant substitution.
\begin{enumerate}
    \item \textbf{Slowly Varying Envelope Demodulation:} We decompose the fast-oscillating discrete carrier wave by introducing a complex-valued slowly varying electric field envelope $E(z,t) \in \mathbb{C}$ such that $x(z,t) = \frac{1}{2}[E(z,t)e^{i\omega t} + \text{c.c.}]$. This operation converts the second-order temporal derivative into a first-order envelope time evolution.
    \item \textbf{Nonlinear Constant Mapping:} The mechanical cubic stiffness coefficient $\beta$ is mapped directly onto the third-order optical Kerr nonlinearity coefficient $\chi^{(3)}$, which governs self-phase modulation inside the high-$Q$ cavities. The mechanical damping $\gamma$ maps onto the baseline linear optical cavity loss $\gamma_{\text{loss}}$.
    \item \textbf{Traveling-Wave Spatiotemporal Modulation:} The discrete phase-shifted inputs are continuously integrated into a macroscale periodic traveling index wave, establishing a phase modulation velocity $v_{\text{mod}} = \Omega/q$.
\end{enumerate}
By executing these steps, the real-valued mother dynamics transform into a dissipative Nonlinear Schr{\"o}dinger Equation (NLSE) architecture, which, upon complex phase rotation, exactly converges to the continuous optical covariant envelope framework:
\begin{equation}
\gamma_{\text{loss}} \frac{\partial E}{\partial t} = -\left( \frac{1}{i}\frac{\partial}{\partial z} - \beta_0 - v_{\text{mod}} \right)^2 E + \mu_{\text{opt}} E - \chi^{(3)} |E|^2 E,
\end{equation}
where the accumulation of dynamic field stress triggers the nucleation of localized optical solitons as autonomous physical interlocks.

\subsection*{4.4 Branch C: Deductive Instantiation of Atomic Toroidal BEC Fields}
To project the mother field onto an ultracold gas platform, specifically a toroidal Bose--Einstein Condensate (BEC), we map the discrete constraints onto macroscopic mean-field superfluid parameters.
\begin{enumerate}
    \item \textbf{Mean-Field Wavefunction Complexification:} The discrete coordinates are upscaled into the macroscopic complex order parameter $\Psi(\theta,t) \in \mathbb{C}$, representing the collective condensation state of alkali atoms within a ring trap of radius $R$.
    \item \textbf{Dissipative Quantum Cooling Mapping:} The discrete damping factor $\gamma$ is mapped onto the phenomenological dissipation coefficient $\gamma_{\text{bec}} \hbar$, modeling the continuous particle and energy exchange with the surrounding thermal cloud.
    \item \textbf{Artificial Gauge Field Generation:} The directional transport term $v_{\text{bias}}$ is physically instantiated by introducing an artificial vector potential $A_{\text{eff}}$ via counter-propagating Raman laser pairs [S15]. This breaks the azimuthal inversion symmetry, inducing an effective rotational velocity $v_{\text{rot}} = A_{\text{eff}}/m$.
    \item \textbf{Interaction Parameter Integration:} The mechanical nonlinearity $\beta$ is replaced by the short-range atomic repulsive interaction strength $g_{1D}$ governed by the s-wave scattering length.
\end{enumerate}
Under the strong dissipative cooling limit where the relaxation dynamics dominate over coherent Schr{\"o}dinger phase rotations, the continuous system reduces precisely to the covariant Dissipative Gross--Pitaevskii Equation (DGPE):
\begin{equation}
\gamma_{\text{bec}} \hbar \frac{\partial \Psi}{\partial t} = -\frac{\hbar^2}{2m} \left( \frac{1}{i}\frac{1}{R}\frac{\partial}{\partial \theta} - \frac{m v_{\text{rot}}}{\hbar} \right)^2 \Psi + \mu_0 \Psi \left( 1 - \frac{g_{1D}}{\mu_0} |\Psi|^2 \right).
\end{equation}
Here, excessive input perturbations break the Landau critical velocity, forcing the system to autonomously generate and eject quantized vortex pairs into the central vacuum, perfectly preserving the information integrity of the computational ring.

\subsection*{4.5 Structural Summary of the Deductive Unification}
By positioning the discrete asymmetric Duffing framework as the universal mathematical mother template, we demonstrate that all three multi-physics platforms are rigorous continuum realizations of a singular underlying topological law. The passive buffering via internal dissipation, the causal filtering via broken spatial symmetry, and the autonomous self-healing via localized defect sequestration are not isolated phenomena of specialized substrates; they are universal geometric assets organically derived from the root discrete mother network.

\clearpage
\section*{Supplementary Note 5:\\
Thermodynamic Derivation of Benchmark Metrics and Comparison with Digital Monitoring
}

In this supplementary note, we outline the microscopic physical models and thermodynamic scaling relations used to derive the comparative performance metrics presented in Table~I of the main text (Appendix~A).

\subsection*{5.1 Intrinsic Relaxation Latency ($\tau_{\text{GL}}$)}
In conventional digital coprocessors, anomaly detection requires digitizing continuous signals via analog-to-digital converters (ADCs), followed by algorithmic variance thresholding executed over multiple clock cycles ($1\text{--}100\ \mu\text{s}$).

In contrast, the response of the superconducting continuous order parameter field $\psi(x,y)$ is governed by the intrinsic Ginzburg--Landau relaxation timescale $\tau_{\text{GL}}$, defined near $T_c$ as [S3]:
\begin{equation}
\tau_{\text{GL}}(T) = \frac{\pi \hbar}{8 k_B (T_c - T)} \approx \frac{\pi \hbar}{8 k_B T_c \left(1 - T/T_c\right)}.
\end{equation}
For pure niobium thin films ($T_c \approx 9.2\text{ K}$) [S9, S10] operating at $T = 4.2\text{ K}$ ($T/T_c \approx 0.45$), the microscopic phase relaxation time evaluates to:
\begin{equation}
\tau_{\text{GL}} \approx \frac{3.1416 \times (1.054 \times 10^{-34}\ \text{J}\cdot\text{s})}{8 \times (1.38 \times 10^{-23}\ \text{J/K}) \times 9.2\text{ K} \times 0.55} \approx 0.59\ \text{ps}.
\end{equation}
Even when accounting for macroscopic vortex transit across the 150-nm track at typical depinning velocities ($v \approx 10^2\text{--}10^3\text{ m/s}$), the spatial reorganization completes within $\tau_{\text{trans}} \approx w / v \approx 0.15\text{--}1.5\text{ ns}$. This represents an operational speedup of four to six orders of magnitude ($10^4\text{--}10^6\times$) relative to digital interrupt routines.

\subsection*{5.2 Energy Consumption of Defect Sequestration}
Active digital error correction in edge processors consumes energy through repetitive CMOS gate switching, where the dynamic energy per clock cycle scales as $E_{\text{digital}} = \alpha C V_{dd}^2$. For a dedicated digital signal processor running an interrupt handler for $10\ \mu\text{s}$ at $100\text{ MHz}$, the minimum energy dissipation is on the order of $\mu\text{J}$ to $\text{mJ}$.

In the continuous superconducting field, defect nucleation and pinning are entirely passive thermodynamic events governed by the Ginzburg--Landau free energy density [S3]. In SI units, the energy cost to nucleate and decouple a pair of quantized vortices from the ring into the central pinning minimum corresponds to the vortex line condensation energy per unit length:
\begin{equation}
\epsilon_v = \frac{\Phi_0^2}{4\pi \mu_0 \lambda^2} \ln\left(\frac{\lambda}{\xi}\right),
\end{equation}
where $\Phi_0 = h/(2e) \approx 2.07 \times 10^{-15}\text{ Wb}$ is the magnetic flux quantum and $\mu_0 = 4\pi \times 10^{-7}\text{ H/m}$ is the vacuum permeability. Multiplying $\epsilon_v$ by the thin-film thickness ($d \approx 50\text{ nm}$) with $\lambda \approx 90\text{ nm}$ and $\xi \approx 40\text{ nm}$ [S9, S10] yields:
\begin{equation}
E_{\text{pair}} \approx 2 \times \epsilon_v \times d \approx 2 \times \left(2.7 \times 10^{-11}\ \text{J/m}\right) \times \left(50 \times 10^{-9}\ \text{m}\right) \approx 2.7 \times 10^{-18}\text{ J} \ (2.7\ \text{aJ}).
\end{equation}
Even when integrating total non-equilibrium viscous dissipation during the entire $t = 5\text{--}15$ perturbation transient ($E_{\text{diss}} = \int v_{\text{total}} dt$), the total dissipated heat remains strictly bounded below $10^{-15}\text{ J}$ (femtojoules). Thus, thermodynamic topological self-healing operates with an energetic efficiency exceeding conventional digital monitors by more than nine orders of magnitude ($> 10^9\times$).

\vspace{2em}
\noindent
\textbf{Supplementary References}
\begin{enumerate}\small
\setlength{\itemsep}{1pt}
\setlength{\parskip}{0pt}
\renewcommand{\labelenumi}{[S\arabic{enumi}]}
\item T. Hikihara and Y. Aoki, ``Autonomous self-repair in coupled Duffing oscillator networks,'' arXiv:2608.26741 (2026).
\item V. L. Ginzburg and L. D. Landau, ``On the theory of superconductivity,'' Zh. Eksp. Teor. Fiz. \textbf{20}, 1064--1082 (1950).
\item M. Tinkham, \textit{Introduction to Superconductivity} (McGraw-Hill, New York, 1975).
\item R. Kato, Y. Enomoto, and S. Maekawa, ``Effects of the surface barrier on a critical state of a type-II superconducting slab,'' Phys. Rev. B \textbf{44}, 6916--6923 (1991).
\item T. Hikihara, Y. Hosoe, and T. Hagiwara, ``Energy flow and causality in coupled non-equilibrium systems,'' in \textit{Proceedings of the 2013 International Symposium on Nonlinear Theory and its Applications (NOLTA 2013)} (2013), pp. 34--37.
\item H. Jaeger, ``The `echo state' approach to analysing and training recurrent neural networks,'' GMD Tech. Rep. \textbf{148}, 13 (2001).
\item G. Tanaka, T. Yamane, J. B. H{\'e}roux, R. Nakane, N. Kanazawa, S. Takeda, H. Numata, D. Nakano, and A. Hirose, ``Recent advances in physical reservoir computing: A review,'' Neural Netw. \textbf{115}, 100--123 (2019).
\item D. R. Hofstadter, ``Energy levels and wave functions of Bloch electrons in rational and irrational magnetic fields,'' Phys. Rev. B \textbf{14}, 2239--2249 (1976).
\item B. W. Maxfield and W. L. McLean, ``Superconducting penetration depth of niobium,'' Phys. Rev. \textbf{139}, A1515--A1522 (1965).
\item A. I. Gubin, K. S. Il'in, S. A. Vitusevich, M. Siegel, and N. Klein, ``Dependence of magnetic penetration depth on the thickness of superconducting Nb thin films,'' Phys. Rev. B \textbf{72}, 064503 (2005).
\item V. R. Manfrinato, L. Zhang, D. Su, H. Duan, R. G. Hobbs, E. A. Stach, and K. K. Berggren, ``Resolution limits of electron-beam lithography toward the atomic scale,'' Nano Lett. \textbf{13}, 1555--1558 (2013).
\item G. S. Oehrlein, ``Reactive ion etching,'' Phys. Today \textbf{39}(10), 26--33 (1986).
\item S. Miyamoto and T. Hikihara, ``Dynamics of fluxoid in a superconducting ring by time-dependent Ginzburg-Landau equation,'' Physica C \textbf{417}, 7--14 (2004).
\item R. Martinenghi, S. Rybalko, M. Jacquot, L. Larger, and Y. K. Chembo, ``Photonic nonlinear transient computing with multiple-delay wavelength dynamics,'' Phys. Rev. Lett. \textbf{108}, 244101 (2012).
\item K. Jim{\'e}nez-Garc{\'i}a, L. J. LeBlanc, R. A. Williams, M. C. Beeler, A. R. Perry, and I. B. Spielman, ``Peierls substitution in an engineered lattice potential,'' Phys. Rev. Lett. \textbf{108}, 225303 (2012).
\end{enumerate}

\end{document}